\documentclass[a4paper]{article}

\usepackage{INTERSPEECH2019}
\usepackage{graphicx}
\usepackage{amsmath,amssymb,caption,array,bm,hyperref}
\usepackage{algorithm,algorithmic,color,multirow,float,subcaption}
\DeclareMathOperator{\vv}{\mathbf{v}}
\title{Multi-modal Sentiment Analysis using Deep Canonical Correlation Analysis}
\name{Zhongkai Sun$^{1*}$, Prathusha K Sarma$^{2*}\thanks{$^*$equal contribution by both authors}$, William Sethares$^3$, Erik P. Bucy$^4$}
\address{
$^{[1,2,3]}$Electrical and Computer Engineering, UW-Madison\\
$^4$CoMC, Texas Tech University}
\email{zsun227@wisc.edu, kameswarasar@wisc.edu,sethares@wisc.edu, erik.bucy@ttu.edu}

\begin{document}

\maketitle
\begin{abstract}
This paper learns multi-modal embeddings from text, audio, and video views/modes of data in order to improve upon downstream sentiment classification. The experimental framework also allows investigation of the relative contributions of the individual views in the final multi-modal embedding. Individual features derived from the three views are combined into a multi-modal embedding using Deep Canonical Correlation Analysis (DCCA) in two ways i) One-Step DCCA and ii) Two-Step DCCA. This paper learns text embeddings using BERT, the current state-of-the-art in text encoders. We posit that this highly optimized algorithm dominates over the contribution of other views, though each view does contribute to the final result. Classification tasks are carried out on two benchmark data sets and on a new Debate Emotion data set, and together these demonstrate that the one-Step DCCA outperforms the current state-of-the-art in learning multi-modal embeddings.

% This paper presents two techniques to obtain multi-modal embeddings from text, audio and video views/modes for sentiment analysis. Multi-modal embeddings are obtained using Deep Canonical Correlation Analysis (DCCA) in a One-Step and Two-Step configuration. The learned multi-modal embeddings are then used in downstream sentiment classification tasks on standard benchmark data sets as well as on a new Debate Emotion data set. Empirical results show that the proposed One-Step and Two-Step DCCA approaches learn the most effective multi-modal embeddings from input data.
\end{abstract}

\section{Introduction}
Various social media platforms make available a variety of multi-modal content generated through expression of opinions and ideologies by social media users in the form of written commentary, podcasts, and lifestyle vlogs on a variety of topics such as politics, entertainment, reviews of movies, products etc. Multi-modal data enables one to understand the interplay of linguistic and behavioral cues, particularly when trying to resolve user sentiment or when studying affective behavior, such as the rise of political populism.

Which is more important in human discourse: text, speech, or video?" We approach this question via an experimental paradigm that solves sentiment classification problems using each feature set (text, audio, video) individually, in pairs, and all three together. This allows us to assess the relative importance of the contribution of each data view/mode\footnote{henceforth we shall use the words view and mode interchangeably}. By necessity, we investigate ways of ``merging'' the feature vectors from the three views. The principal result is that more views give better classification, though there is an asymmetry in the development and quality of algorithms for extracting the three views that likely biases any quantitative interpretation of these results.

Recent work on multi-modal~\cite{poria2018multimodal},~\cite{hu2018multimodal} and multi-view~\cite{arora2013multi} sentiment analysis combine text, speech and video/image as distinct data views from a single data set. The idea is to make use of written language along with voice modulation and facial features either by encoding for each view individually and then combining all three views as a single feature~\cite{poria2018multimodal},~\cite{hu2018multimodal} or by learning correlations between views and then combining them in a correlated space~\cite{arora2013multi}.
Each technique has demonstrated significant improvements in classification accuracy when used to detect sentiment. In addition to improving upon performance metrics for a downstream task such as classification, multi-modal data also enables one to study which view (or combination of views) is most efficient in understanding user behavior. For example, when studying populism~\cite{bucy2018performing, Joo2019auto}, visual and tonal expressions of rage have been found to be key characteristics of populist behavior. Since there is a rising interest in using multimodal data for tasks other than sentiment analysis~\cite{hu2018multimodal}, it is important to explore how and to what extent each individual view contributes towards the overall success on a downstream task for a particular data set.

This paper makes the following contributions: i) Learn multi-modal data embeddings using Deep Canonical Correlation Analysis in a One-Step and Two-Step framework to combine text, audio and video views for the improvement of sentiment/emotion detection. The Two-Step DCCA framework further helps to explore the interplay between audio, video and text features when learning multi-modal embeddings for classification tasks. ii) Encode text using BERT~\cite{devlin2018bert}, the current state-of-the-art in text encoders to obtain fixed-length representations for text. There is little literature that uses pre-trained BERT encoders as features without additional fine-tuning. This work adds to the growing body of work that applies BERT as a pre-fixed feature and iii) perform empirical evaluations on benchmark data sets such as CMU-MOSI~\cite{zadeh2016mosi} and CMU-MOSEI~\cite{cmumoseiacl2018} along with a new Debate Emotion data set introduced by~\cite{bucy2018performing}.%\footnote{add about making data available}.

The rest of the paper is organized as follows, Section~\ref{relwork} presents related work and Section~\ref{methods}
describes the methodology used in the paper. Section~\ref{exps} presents results and Section~\ref{conc} concludes.

% \textcolor{blue}{Additional motivations based on results}

% In this paper, we make use of Deep Canonical Correlation Analysis (DCCA) to learn correlated projections between the textual, acoustical and visual views of a given data set. Text is encoded using BERT~\cite{devlin2018bert} a transformer model that encodes text by making use of conditional contexts. Audio is encoded using \textcolor{blue}{name method} and video is encoded using \textcolor{blue}{name video method}.

\section{Related Work}\label{relwork}

The idea of combining multi-modal text, audio and video features expressed in this paper is closest to that of \cite{poria2018multimodal} which encodes text, speech, and visual signals using using a BiLSTM encoder, openSMILE, and 3D-CNN respectively. Encoded outputs are then concatenated and passed through a classifier. In contrast, this paper employs multi-modal embeddings that are obtained by learning correlated representations of text, audio, and video views using Deep Canonical Correlation Analysis (DCCA)~\cite{andrew2013deep} as in~\cite{dumpalaaudio}. This approach is similar to the use of Generalized Canonical Correlation Analysis (GCCA) as in~\cite{rastogi2015multiview}. Recent work in text based sentiment analysis \cite{sarma2018domain}--\cite{benton2016learning} has demonstrated the effectiveness of statistical methods like Canonical Correlation Analysis (CCA) and its variants such as GCCA and DCCA on various uni-, bi-, and multi-modal learning tasks.

\section{Methods}\label{methods}

This section briefly reviews Deep Canonical Correlation Analysis (DCCA) and outlines the methods used to obtain unimodal features. This section also outlines the procedure used to obtain the multi-modal embeddings used for experiments in Section~\ref{exps}.

\subsection{Deep Canonical Correlation Analysis (DCCA)}

Classic Canonical Correlation Analysis (CCA)~\cite{hotelling1992relations} is a statistical technique used to find a linear subspace in which two sets of random variables with finite second moments are maximally correlated. This idea is applied in the context of multi-modal learning by considering each view/modality of the data to be a random variable, and then using CCA to find the subspace such that non-discriminative features in each view are largely un-correlated, and hence can be filtered out. The natural generalization of this idea, to learning the subspace via non-linear projections obtained from feed forward neural networks, is called Deep CCA. 

A DCCA network has input $(x_{1},x_{2})$, which denotes two input views (corresponding to the same input). Let 
\[ f_{i}(x_{i};\theta_{i}) = s_{i}(W_{d}h_{d-1}+b_{d}) \]
denote the final layer of a $d$ layered neural network, whose first layer is $h_{i}=s_{i}(W_{i}x_{i}+b_{i})$ with input $x_{i}$. Thus $f_{1}(x_{1};\theta_{1})$ and $f_{2}(x_{2};\theta_{2})$ represent two neural networks used to encode the two views $(x_{1},x_{2})$ of the data and are parameterized by $\theta_{1}=(W_{1,d},b_{1,d})$ and $\theta_{2}=(W_{2,d},b_{2,d})$. The objective of DCCA is to determine the parameters of the two networks such that
\begin{align}
(\theta_{1}^{*},\theta_{2}^{*})=\text{argmax}_{\theta_{1},\theta_{2}}\text{corr}(f_{1}(x_{1};\theta_{1}),f_{2}(x_{2};\theta_{2}))
\end{align}
where $\text{corr}$ denotes the statistical correlation between $x_{1}$ and $x_{2}$.

\subsection{Unimodal Feature Extraction}
\begin{itemize}
  \item \textbf{Text Encoding:} To encode text in the data, we use pre-trained BERT (Bidirectional Encoder Representations from Transformers). Like the name suggests, BERT is a transformer based language model that conditions jointly on the left and right of a given word. Typically, the BERT encoder is fine-tuned to a particular task by learning an additional task-specific weight layer. We use the output from the BERT encoder, pre-trained on a large corpus of Wikipedia+Book corpus data, and do not perform additional fine-tuning. The choice of encoder is motivated by the success of BERT in achieving the state-of-the-art in several NLP tasks such as sentiment analysis, question-answering, textual entailment etc. Input text in Section~\ref{exps} is encoded using BERT-base, and all text embeddings are of size 768. Henceforth, denote embedded text by $\vv_{t}$.
  \item \textbf{Audio Encoding:} Audio features from data are extracted using COVAREP~\cite{degottex2014covarep}, that extracts MFCC, pitch, peak slope, and other acoustic features from audio frames. Audio embedding for each video is the the average of the audio vectors extracted from each individual frame feature. Resultant audio embeddings are of size 74. Henceforth, audio embeddings are denoted by $\vv_{a}$.
  \item \textbf{Video Encoding:} Framewise features from video stream are extracted using a combination of FACET\footnote{\url{https://imotions.com/facial-expressions/}} and OpenFace 2.0~\cite{baltrusaitis2018openface}\footnote{\url{https://github.com/TadasBaltrusaitis/OpenFace}}. For each 10 second duration video, video-level feature vectors are obtained by averaging across the feature vectors corresponding to individual frames. Video embeddings are denoted by $\vv_{v}$. For two (MOSI and MOSEI) of the three data sets considered in Section~\ref{exps} video features are extracted by using FACET. On the Debate Emotion data set, video features of 2016 debate videos are represented as facial action units features as in~\cite{baltruvsaitis2015cross}. Resultant video embeddings are of size 35. Henceforth, video embeddings are denoted by $\vv_{v}$.
\end{itemize}

\subsection{Methodology}
DCCA accepts two views of data at a time and learns a correlated subspace. Since we are working with three views of data, all three views must be combined. We consider two different procedures. The One-Step DCCA concatenates the audio and video features and applies DCCA to this combined audio-video feature and the text features. The Two-Step DCCA combines two of the views in the first step, and then combines the third with the first two in its second step. These are briefly explained in the following sections.

\subsubsection{One-Step DCCA}

In this set up, one input view to DCCA is fixed to be text encoded ($\vv_{t}$) by BERT and the other input view to the DCCA algorithm is a concatenation of the audio and video embeddings ($\vv_{a}|\vv_{v}$). The intuition behind this is that, most often written/transcribed text involves an explicit statement of sentiment while voice modulation and facial features may convey less explicit though perhaps more emotion-laden information. For example, in the debate data~\cite{bucy2018performing} some speakers are more controlled in their speech (as they express aggression through carefully planned statements) as opposed to other speakers who have explicit vocal and tonal features marking their aggression.

Algorithm~\ref{algo1} describes the One-Step DCCA algorithm. After applying DCCA once to obtain correlated representations of text ($\bar{\vv}_{t}$) and audio-video ($\bar{\vv}_{a,v}$), the input embeddings are concatenated with the correlated embeddings to obtain the final representation of the text and audio-video views as indicated in line 3. The final multi-view embedding $\vv_{\text{multi}}$ is obtained by concatenating the final text and audio-video representations.
\begin{algorithm}
 \begin{algorithmic}[1]
 \REQUIRE $\vv_{t},\vv_{a},\vv_{v}$
 \STATE Initialize $\vv_{1} = \vv_{t}$, $\vv_{2}=[\vv_{a}|\vv_{v}]$.
 \STATE $(\bar{\vv}_{1}, \bar{\vv}_{2}) \leftarrow \text{DCCA}(\vv_{1},\vv_{2})$.
 \STATE $\hat{\vv}_{t} = [\bar{\vv}_{1}|\vv_{1}]$, $\hat{\vv}_{a,v} = [\bar{\vv}_{2}|\vv_{2}]$.
 \STATE Return $\vv_{\text{multi}} = [\hat{\vv}_{t}|\hat{\vv}_{a,t}]$.
 \end{algorithmic}
\caption{One-Step DCCA\label{algo1}}
\end{algorithm}

\subsubsection{Two-Step DCCA}

Since we are interested in studying the interplay of audio, video, and text views when learning multi-modal embeddings, in this framework we empirically explore the optimal combination of input views to DCCA. For example, we can fix one input view to be audio. The second view is obtained as the output from a separate DCCA operation that takes as inputs text and video views. The correlated text and video views are then concatenated to form second input view. This way we perform two DCCA steps as suggested by the name of the method. 

Algorithm~\ref{algo2} briefly describes the Two-Step DCCA algorithm. This algorithm is the same as Algorithm~\ref{algo1} with the exception of lines 3, where we take the output of the first DCCA step to obtain one input view for the second DCCA step. We posit that unlike the One-Step DCCA, the Two-Step DCCA would ideally perform better since we correlate two views first, before correlating the third view. However as explained via the results in Table~\ref{algo2} due to the large variation in dimension across the three views, we do not see the expected improvements over One-Step DCCA.

\begin{algorithm}
 \begin{algorithmic}[1]
 \REQUIRE $\vv_{t},\vv_{a},\vv_{v}$
 \STATE Initialize $(\vv_{1} = \vv_{t}$, $\vv_{2}=\vv_{v})$ or $(\vv_{1}=\vv_{v},\vv_{2}=\vv_{a})$ or $(\vv_{1}=\vv_{t},\vv_{2}=\vv_{a})$.
 \STATE $(\bar{\vv}_{1}, \bar{\vv}_{2}) \leftarrow \text{DCCA}(\vv_{1},\vv_{2})$.
 \STATE Initialize $\vv'_{1} = [\bar{\vv}_{1}|\bar{\vv}_{2}]$ and $\vv'_{2} = \vv_{a} or \vv_{t} or \vv_{v}$.
 \STATE $(\bar{\vv}'_{1}, \bar{\vv}'_{2}) \leftarrow \text{DCCA}(\vv'_{1},\vv'_{2})$.
 \STATE $\hat{\vv}'_{1} = [\bar{\vv}'_{1}|\vv'_{1}]$, $\hat{\vv}'_{2}=[\bar{\vv}'_{2}|\vv'_{2}]$.
 \STATE Return $\vv_{\text{multi}} = [\hat{\vv}'_{1}|\hat{\vv}'_{2}]$.
 \end{algorithmic}
\caption{Two-Step DCCA\label{algo2}}

\end{algorithm}

\subsubsection{Sentiment Classification}
Multi-view embeddings obtained from One-Step DCCA and Two-Step DCCA are input to a logistic regression classifier to predict the sentiment label for test data sets in Section~\ref{exps}.
\begin{table*}
\centering
\caption{This table presents accuracy and F-score for One-Step DCCA and baseline methods on MOSI, MOSEI and Debate Emotion data sets. Best performing algorithm and modality are indicated in boldface. Star marked results correspond to numbers reported in original publication.}
\begin{tabular}{|l|l|l|l|l|l|l|}
\hline
\multicolumn{1}{|c|}{Data View}&\multicolumn{2}{|c|}{Debate Emotion}&\multicolumn{2}{|c|}{CMU-MOSI}&\multicolumn{2}{|c|}{CMU-MOSEI}\\
\cline{2-7}&Acc&F-score&Acc&F-score&Acc&F-score\\
\hline
Audio&85.0&85.8&44.5&45.0&51.21&51.94\\
Video&81.0&82.06&44.0&44.5&58.75&59.23\\
Text&77.5&76.8&78.8&79.17&80.23&83.00\\
Audio+Video&82.5&83.0&49.0&50.1&62.46&63.03\\
Audio+Text&85.0&85.3&79.8&79.7&82.88&83.2\\
Video+Text&85.5&83.2&79.44&79.41&83.05&83.12\\
\textbf{Audio+Video+Text (One-Step DCCA)}&\textbf{93.0}&\textbf{93.1}&\textbf{80.6}&\textbf{80.57}&\textbf{83.62}&\textbf{83.75}\\
Audio+Video+Text (GCCA)&88.0&87.9&78.0&77.36&83.02&81.16\\
Audio+Video+Text (Logistic Reg)&91.0&90.9&79.5&76.6&82.97&83.20\\
Audio+Video+Text (bc-LSTM+3D-CNN+openSMILE)&N/A&N/A&78.8*&N/A&N/A&N/A\\
Audio+Video+Text(Graph Memory Fusion Network)&N/A&N/A&N/A*&N/A*&76.9* & 77.0*\\
\hline
% &1&1&1&1\\
% \hline
% \multicolumn{2}{|c|}{Audio}&-&-&-\\
% \hline
\end{tabular}
\label{exps1}
\end{table*}

% \begin{table*}
% \caption{This table presents accuracy from One-Step DCCA and baseline methods on MOSI, MOSEI and Debate Emotion data sets. Best performing algorithm and modality are indicated in boldface. Star marked results correspond to numbers reported in original publication.}
% \centering
% \resizebox{0.7\textwidth}{!}{\begin{tabular}{|l|l|l|l|}
% % \begin{tabular}{|l|l|l|l|}
% \hline
% Data View &Acc on Debate& Acc on MOSI&Acc on MOSEI\\
% \hline
% Audio&85&44.5&58.0\\
% Video&81&44&57.6\\
% Text&77.5&78.8&80.9\\
% Audio+Video&82.5&49&59.5\\
% Audio+Text&85&79.8&0\\
% Video+Text&85.5&78.6&0\\
% \textbf{Audio+Video+Text (One-Step DCCA)}&\textbf{93}&\textbf{80.6}&\textbf{81.28}\\
% Audio+Video+Text (GCCA)&88&61.5&75.21\\
% Audio+Video+Text (Logistic Reg)&91&79.4&80.61\\
% Audio+Video+Text (bc-LSTM+3D-CNN+openSMILE)&N/A&78.8*&N/A\\
% \hline
% \end{tabular}}

% \label{exps1}
% \end{table*}
% i)One-Step DCCA: Here we input to DCCA text encoded from BERT as one view and a concatenation of audio and video encodings as the other view. ii)Two-Step DCCA: As the name suggest we perform two DCCA step, the first takes as input audio and video (nips paper) and learns a correlated subspace for the audio and video inputs. The output from the first step DCCA is then concatenated.

\section{Experiments}\label{exps}
This section first describes the different test data sets used and the baseline methods that are evaluated against embeddings obtained from One-Step DCCA and Two-Step DCCA. Multi-modal embeddings obtained from DCCA methods are input to a logistic regression classifier and accuracy and F-scores on test data sets are reported as the performance metrics.
\subsection{Test Data Sets}
The following data sets are used for testing,
\begin{itemize}
  \item \textbf{CMU-MOSI:} This is a standard benchmark data set of product reviews curated by 93 users and introduced by~\cite{zadeh2016mosi}. Reviews are in the form of videos that are segmented into clips. Each clip is assigned a sentiment score between -3 to 3 by five annotators. Sentiment scores are further binarized as `positive' and `negative', by assigning all reviews having scores $\geq 0$ as `positive' and all scores $<0$ as `negative'. There are a total of 2198 data points in the classification task. Predetermined splits of training data (1283 points), validation data (229 points) and test data (686 points) are used in our experiments.

  \item \textbf{CMU-MOSEI:} This data set~\cite{cmumoseiacl2018} is similar to MOSI and is also annotated at the utterance/clip level. Each utterance is assigned a sentiment score between -3 to 3. Frames scored $(0,3]$ are labeled as `positive' and scores between $[-3,0)$  are labeled as `negative'. There area total of 17859 data points available for binary sentiment classification. Data is partitioned into predetermined train (12787 points), validation (3634 points) and test (1438 points) splits. Raw features for CMU-MOSI and CMU-MOSEI are obtained from CMU-Multimodal SDK~\cite{zadeh2018multi} .

  \item \textbf{Debate Emotion:} This data set was curated by~\cite{bucy2018performing} by combining the first and third of the 2016 presidential debates. Data is divided into short videos each of a 10 second duration. Videos that contained multiple speakers in a 10-second duration were not considered, thereby restricting each video to a single speaker. This results in a total of 800 short videos, with a single candidate speaking for 10 seconds. Candidate videos were annotated for `aggression' based on candidate expression on three views: verbal, facial and tonal. An aggression label of 1 was assigned to the video if the candidates expressed anger or aggression in either of the three views. This data set consists of 800 data points, partitioned into train (510 points), validation (90 points) and test (200 points) splits.

  % \textcolor{blue}{This dataset has been split into three parts, train(510), test(200), and validation(90)}For additional details regarding data collection and annotation we refer readers to~\cite{bucy2018performing}. 
\end{itemize}

\subsection{Baseline Algorithms}
Since the focus of these experiments is to demonstrate i) the combination of three views is better than unimodal feature embeddings in sentiment analysis tasks and ii) study the contribution of various views in multi-modal embeddings used for classification we compare against the following baselines in this work, 
\begin{itemize}
	\item \textbf{Unimodal features:} To empirically confirm our hypothesis that three views do better than one, multi-modal embeddings are compared against text, audio and video embeddings when input separately to a logistic regression classifier. Additionally we also compare against bi-modal features obtained by taking concatenations of audio/video, video/text, text/audio and passing the concatenated features as input to a logistic regression classifier.
	\item \textbf{Generalized Canonical Correlation Analysis:} GCCA~\cite{gcca} aims to find a correlation subspace in which weighted combinations of all the input views are correlated. Since each view may contribute differently towards a downstream task, GCCA employs a technique to learn weights corresponding to the importance of each view. A discussion detailing the mechanics of the algorithm are beyond the scope of this paper and we refer readers to~\cite{benton2016learning} for the same. Since we explore a Two-Step approach as a potential combination technique for the three views, it is best compared against a technique like GCCA, that learns a weighted combination of all views.
	\item \textbf{bc-LSTM+3D-CNN+openSMILE:} This algorithm was introduced by~\cite{poria2018multimodal} and uses a bidirectional LSTM to encode for text, 3D-CNN and openSMILE to obtain visual and audio embeddings. 
	\item \textbf{Graph Memory Fusion Network:} This algorithm~\cite{cmumoseiacl2018} proposes a dynamic fusion graph to analyze the interactions between different modalities at the word level. LSTMs are used to  capture information for the whole sequence.
\end{itemize}

\textbf{Hyperparameters for DCCA:} DCCA implementation in this work follows that of~\cite{andrew2013deep} with three fully connected feed-forward layers and a ReLU activation applied to the output of each connected layer. DCCA objective is optimized using RMSProp as in the original implementation. Sizes of connected layers are determined via grid search. A similar technique is used to determine parameters of the Logistic Regression classifier.

\subsection{Experimental Results}
\begin{table}
\centering
\caption{This table presents results from the Two-Step DCCA procedure for all combination of input views. Accuracy of the best performing combination on each data set is represented in boldface.}
% \resizebox{0.65\textwidth}{!}{\begin{tabular}{|l|l|l|l|}
\resizebox{\columnwidth}{!}{\begin{tabular}{|l|l|l|l|l|l|l|}

\hline
\multicolumn{1}{|c|}{Data View}&\multicolumn{2}{|c|}{Debate Emotion}&\multicolumn{2}{|c|}{CMU-MOSI}&\multicolumn{2}{|c|}{CMU-MOSEI}\\
\cline{2-7}&Acc&F-score&Acc&F-score&Acc&F-score\\
% Data View combination &Acc on Debate& Acc on MOSI&Acc on MOSEI\\
\hline
$\vv_{1}=\vv_{a}$,$\vv_{2}=\vv_{v}$,$\vv'_{2}=\vv_{t}$&\textbf{92.5}&\textbf{92.57}&\textbf{80.17}&\textbf{80.32}&\textbf{83.57}&\textbf{83.71}\\
$\vv_{1}=\vv_{t}$,$\vv_{2}=\vv_{v}$,$\vv'_{2}=\vv_{a}$&92.0&92.05&79.33&79.46&83.23&83.30\\
$\vv_{1}=\vv_{t}$,$\vv_{2}=\vv_{a}$,$\vv'_{2}=\vv_{v}$&91.5&91.34&79.45&79.44&83.19&83.33\\
\hline
\end{tabular}}
\label{exps2}
\end{table}

Table~\ref{exps1} presents accuracy and F-score obtained by baseline methods and One-Step DCCA on the MOSI, MOSEI and Debate Emotion data sets. Results from Table~\ref{exps1} indicate that, unsurprisingly, making use of all three views when learning feature embeddings provides the best result. Furthermore, learning combinations in a correlated space using DCCA consistently outperforms other baselines on all three data sets. Note that, the performances reported for the bc-LSTM+3D-CNN+openSMILE baseline and the Graph Memory Fusion Network are as reported in~\cite{poria2018multimodal} and~\cite{cmumoseiacl2018}. We do not reproduce code from~\cite{poria2018multimodal},~\cite{cmumoseiacl2018} for evaluation on Debate Emotion dataset. Besides, these two baselines do not directly compare against our method, since they operate at the word level with different text features. However, since these algorithms achieve the state-of-the-art on the MOSEI and MOSI data set, they are included in this work.
All other baselines were reproduced on the test data sets using parameters reported in their original implementations.
% \footnote{There is considerable difference in the features used for learning multi-modal embeddings with DCCA and the features used in~\cite{cmumoseiacl2018}. Comparisons are only between reported performance metrics.\textcolor{red}{I think we can delete this footnode?}}. 

Table~\ref{exps2} represents results from the Two-Step DCCA process. From the emipirical results it can be noted that i) Two-Step DCCA performs just as well as One-Step DCCA and not better and ii) perfoming DCCA first with audio and video as input views and then with text as input view to the second DCCA step is the best performing combination. Note that the dimension of the multi-modal embedding learned using DCCA is upper bounded by $\leq \text{min}(d_{1},d_{2})$ where, $d_{1}$ is the dimension of first input view and $d_{2}$ is the dimension of the second input view. Since there is a large disparity in the dimensions of the text (768) and audio (74) and video (35) views, performing Two-Step DCCA results in multi-modal embedding that is smaller in size that the multi-modal embedding obtained from One-Step DCCA. This loss in dimensionality may lead to information loss further leading to lower quality multi-modal embeddings. The second result is not too surprising either. Given that the text embeddings are highly optimized when compared to the audio and video embeddings, it is not surprising that learning correlations between audio and video embeddings and then combining them with a text embedding produces the best result.
% \begin{table}
% \resizebox{\columnwidth}{!}{\begin{tabular}{|l|l|l|l|l|l|l|}
% \hline
% Data View on &angry & happy &sad &fear & disgust & surprise\\
% \hline
% Audio& 62.3 & 60.5 &58.31 &57.63 &64.24 &52.28\\
% Video& 56.1 & 66.58&58.77 &54.43 &60.26&55.78\\
% Text& 61.4 & 64.56 &60.33 &60.3 &67.8&62.29\\
% Audio+Video& 66.64&49&60.87 &59.58 &64.77&55.12\\
% Audio+Text& 63.51&0&0&0&0&0\\
% Video+Text& 63.2&0&0&0&0&0\\
% Audio+Video+Text (One-step DCCA)&67.01&69.18&63.17&63.75&69.70&65.08\\
% Audio+Video+Text (GCCA)&88&61.5&0&0&0&\\
% Audio+Video+Text (Logistic Reg)&66.58&66.64&61.51&62.14&69.37&63.82\\
% \hline
% \end{tabular}}
% \caption{}
% \label{exps1}
% \end{table}

% Authors must proofread their PDF file prior to submission to ensure it is correct. Authors should not rely on proofreading the Word file. Please proofread the PDF file before it is submitted.
\section{Discussions and Conclusions}\label{conc}

The key issue is this: which is more important, text, speech, or video? Being able to process all three views of human discourse simultaneously allows consideration of the basic question of the relative contributions of the semantics, the spoken delivery, and the accompanying images. Conventional wisdom would be that the text (the ``meaning'') of a statement is the most significant. Common sense argues that the spoken word can have an impact -- one can usually distinguish an argument from a casual conversation even in an unknown language. Erik Bucy~\cite{grabe2009image} argues that it is the images that can be the most significant aspect in the political framing of issues. Our experiments speak to this issue in a concrete way by showing the classification accuracy using two of the views/modes improves on any single view, and that using all three results in the greatest improvement. Digging deeper, the experiments suggest better (and worse) ways of structuring the interactions between the modalities, which we explore by contrasting the ``one-step" and ``two-step approaches". 

At this point, conclusions about the relative importance of the three views should not be taken too quantitatively. First, algorithms such as BERT designed for deriving text-based features are quite sophisticated and have been trained on Wikipedia-sized corpuses. In contrast, both audio and video feature extraction has not received nearly the attention that has been given to text. Second, the corpuses on which the speech and video have been studied are comparatively small. This means that the results likely underestimate the importance of these two views and accentuate the importance of the text. Third, we have no way of knowing if our chosen classification tasks are representative of other common tasks. Results seem overall analogous across three data sets and two tasks, but this is far from a generic representation. Finally, we have no reason to believe that our method of merging the views is optimal. Indeed, there are many possibilities, and while the trends tend to be similar, there is a lot of variation as different (hyper)parameters are considered and different structures are investigated.

% \section{Acknowledgements}
% \textcolor{blue}{acknowledge SMAD members, add bibtex for NIPS paper.}
% The ISCA Board would like to thank the organizing committees of the past INTERSPEECH conferences for their help and for kindly providing the template files. \\
% Note to authors: Authors should not use logos in acknowledgement section; rather authors should acknowledge corporations by naming them only.

\bibliographystyle{IEEEtran}

\bibliography{refbib}

\begin{thebibliography}{10}
\providecommand{\url}[1]{#1}
\csname url@samestyle\endcsname
\providecommand{\newblock}{\relax}
\providecommand{\bibinfo}[2]{#2}
\providecommand{\BIBentrySTDinterwordspacing}{\spaceskip=0pt\relax}
\providecommand{\BIBentryALTinterwordstretchfactor}{4}
\providecommand{\BIBentryALTinterwordspacing}{\spaceskip=\fontdimen2\font plus
\BIBentryALTinterwordstretchfactor\fontdimen3\font minus
  \fontdimen4\font\relax}
\providecommand{\BIBforeignlanguage}[2]{{%
\expandafter\ifx\csname l@#1\endcsname\relax
\typeout{** WARNING: IEEEtran.bst: No hyphenation pattern has been}%
\typeout{** loaded for the language `#1'. Using the pattern for}%
\typeout{** the default language instead.}%
\else
\language=\csname l@#1\endcsname
\fi
#2}}
\providecommand{\BIBdecl}{\relax}
\BIBdecl

\bibitem{poria2018multimodal}
S.~Poria, N.~Majumder, D.~Hazarika, E.~Cambria, A.~Gelbukh, and A.~Hussain,
  ``Multimodal sentiment analysis: Addressing key issues and setting up the
  baselines,'' \emph{IEEE Intelligent Systems}, vol.~33, no.~6, pp. 17--25,
  2018.

\bibitem{hu2018multimodal}
A.~Hu and S.~Flaxman, ``Multimodal sentiment analysis to explore the structure
  of emotions,'' in \emph{Proceedings of the 24th ACM SIGKDD International
  Conference on Knowledge Discovery \& Data Mining}.\hskip 1em plus 0.5em minus
  0.4em\relax ACM, 2018, pp. 350--358.

\bibitem{arora2013multi}
R.~Arora and K.~Livescu, ``Multi-view cca-based acoustic features for phonetic
  recognition across speakers and domains,'' in \emph{2013 IEEE International
  Conference on Acoustics, Speech and Signal Processing}.\hskip 1em plus 0.5em
  minus 0.4em\relax IEEE, 2013, pp. 7135--7139.

\bibitem{bucy2018performing}
E.~P. Bucy, J.~M. Foley, J.~Lukito, L.~Doroshenko, D.~V. Shah, J.~Pevehouse,
  C.~Wells, and E.~P. Bucy, ``Performing populism: Trump’s transgressive
  debate style and the dynamics of twitter response,'' \emph{New Media \&
  Society}, 2018.

\bibitem{Joo2019auto}
J.~Joo, E.~P. Bucy, and C.~Seidel, ``Automated coding of televised leader
  displays: Detecting nonverbal political behavior with computer vision,'' in
  \emph{International Journal of Communication}, In press.

\bibitem{devlin2018bert}
J.~Devlin, M.-W. Chang, K.~Lee, and K.~Toutanova, ``Bert: Pre-training of deep
  bidirectional transformers for language understanding,'' \emph{arXiv preprint
  arXiv:1810.04805}, 2018.

\bibitem{zadeh2016mosi}
A.~Zadeh, R.~Zellers, E.~Pincus, and L.-P. Morency, ``Mosi: multimodal corpus
  of sentiment intensity and subjectivity analysis in online opinion videos,''
  \emph{arXiv preprint arXiv:1606.06259}, 2016.

\bibitem{cmumoseiacl2018}
A.~Zadeh, P.~P. Liang, J.~Vanbriesen, S.~Poria, E.~Cambria, M.~Chen, and L.-P.
  Morency, ``Multimodal language analysis in the wild: {CMU-MOSEI} dataset and
  interpretable dynamic fusion graph,'' in \emph{Association for Computational
  Linguistics (ACL)}, 2018.

\bibitem{andrew2013deep}
G.~Andrew, R.~Arora, J.~Bilmes, and K.~Livescu, ``Deep canonical correlation
  analysis,'' in \emph{International conference on machine learning}, 2013, pp.
  1247--1255.

\bibitem{dumpalaaudio}
S.~H. Dumpala, I.~Sheikh, R.~Chakraborty, and S.~K. Kopparapu, ``Audio-visual
  fusion for sentiment classification using cross-modal autoencoder.''

\bibitem{rastogi2015multiview}
P.~Rastogi, B.~Van~Durme, and R.~Arora, ``Multiview lsa: Representation
  learning via generalized cca,'' in \emph{Proceedings of the 2015 Conference
  of the North American Chapter of the Association for Computational
  Linguistics: Human Language Technologies}, 2015, pp. 556--566.

\bibitem{sarma2018domain}
P.~K. Sarma, Y.~Liang, and B.~Sethares, ``Domain adapted word embeddings for
  improved sentiment classification,'' in \emph{Proceedings of the 56th Annual
  Meeting of the Association for Computational Linguistics (Volume 2: Short
  Papers)}, vol.~2, 2018, pp. 37--42.

\bibitem{benton2016learning}
A.~Benton, R.~Arora, and M.~Dredze, ``Learning multiview embeddings of twitter
  users,'' in \emph{Proceedings of the 54th Annual Meeting of the Association
  for Computational Linguistics (Volume 2: Short Papers)}, vol.~2, 2016, pp.
  14--19.

\bibitem{hotelling1992relations}
H.~Hotelling, ``Relations between two sets of variates,'' in
  \emph{Breakthroughs in statistics}.\hskip 1em plus 0.5em minus 0.4em\relax
  Springer, 1992, pp. 162--190.

\bibitem{degottex2014covarep}
G.~Degottex, J.~Kane, T.~Drugman, T.~Raitio, and S.~Scherer, ``Covarep—a
  collaborative voice analysis repository for speech technologies,'' in
  \emph{2014 ieee international conference on acoustics, speech and signal
  processing (icassp)}.\hskip 1em plus 0.5em minus 0.4em\relax IEEE, 2014, pp.
  960--964.

\bibitem{baltrusaitis2018openface}
T.~Baltrusaitis, A.~Zadeh, Y.~C. Lim, and L.-P. Morency, ``Openface 2.0: Facial
  behavior analysis toolkit,'' in \emph{2018 13th IEEE International Conference
  on Automatic Face \& Gesture Recognition (FG 2018)}.\hskip 1em plus 0.5em
  minus 0.4em\relax IEEE, 2018, pp. 59--66.

\bibitem{baltruvsaitis2015cross}
T.~Baltru{\v{s}}aitis, M.~Mahmoud, and P.~Robinson, ``Cross-dataset learning
  and person-specific normalisation for automatic action unit detection,'' in
  \emph{2015 11th IEEE International Conference and Workshops on Automatic Face
  and Gesture Recognition (FG)}, vol.~6.\hskip 1em plus 0.5em minus 0.4em\relax
  IEEE, 2015, pp. 1--6.

\bibitem{zadeh2018multi}
A.~Zadeh, P.~P. Liang, S.~Poria, P.~Vij, E.~Cambria, and L.-P. Morency,
  ``Multi-attention recurrent network for human communication comprehension,''
  \emph{arXiv preprint arXiv:1802.00923}, 2018.

\bibitem{gcca}
P.~Horst, ``Generalized canonical correlations and their applications to
  experimental data,'' in \emph{Journal of Clinical Psychology, 17(4)}, 1961.

\bibitem{grabe2009image}
M.~E. Grabe and E.~P. Bucy, \emph{Image bite politics: News and the visual
  framing of elections}.\hskip 1em plus 0.5em minus 0.4em\relax Oxford
  University Press, 2009.

\end{thebibliography}

% \begin{thebibliography}{9}
% \bibitem[1]{Davis80-COP}
%   S.\ B.\ Davis and P.\ Mermelstein,
%   ``Comparison of parametric representation for monosyllabic word recognition in continuously spoken sentences,''
%   \textit{IEEE Transactions on Acoustics, Speech and Signal Processing}, vol.~28, no.~4, pp.~357--366, 1980.
% \bibitem[2]{Rabiner89-ATO}
%   L.\ R.\ Rabiner,
%   ``A tutorial on hidden Markov models and selected applications in speech recognition,''
%   \textit{Proceedings of the IEEE}, vol.~77, no.~2, pp.~257-286, 1989.
% \bibitem[3]{Hastie09-TEO}
%   T.\ Hastie, R.\ Tibshirani, and J.\ Friedman,
%   \textit{The Elements of Statistical Learning -- Data Mining, Inference, and Prediction}.
%   New York: Springer, 2009.
% \bibitem[4]{YourName17-XXX}
%   F.\ Lastname1, F.\ Lastname2, and F.\ Lastname3,
%   ``Title of your INTERSPEECH 2019 publication,''
%   in \textit{Interspeech 2019 -- 20\textsuperscript{th} Annual Conference of the International Speech Communication Association, September 15-19, Graz, Austria, Proceedings, Proceedings}, 2019, pp.~100--104.
% \end{thebibliography}

\end{document}